\documentclass[prl,aps,twocolumn, floatfix, superscriptaddress,showpacs]{revtex4}
\usepackage{amsmath,bm,graphicx}
\usepackage{amssymb}
\usepackage{amsfonts}

\usepackage{graphicx}
\usepackage{hyperref}
\usepackage{latexsym}

\def\be{\begin{equation}}
\def\ee{\end{equation}}
\def\bea{\begin{eqnarray}}
\def\eea{\end{eqnarray}}

\begin{document}
\title{Catch bond mechanism in Dynein motor driven collective transport}
\author{Anil Nair}
\affiliation{Department of Physics, Savitribai Phule Pune University, Ganeshkhind, Pune 411007, India}
\author{Sameep Chandel}
\affiliation{Indian Institute of Science Education and Research Mohali, Knowledge City, Punjab 140306, India}
\author{Mithun K Mitra}
\affiliation{Department of Physics, IIT Bombay, India}
\author{Sudipto Muhuri}
\affiliation{Department of Physics, Savitribai Phule Pune University, Ganeshkhind, Pune 411007, India}
\author{Abhishek Chaudhuri}
\affiliation{Indian Institute of Science Education and Research Mohali, Knowledge City, Punjab 140306, India}

\begin{abstract}
Recent experiments have demonstrated that dynein motor exhibits {\it catch bonding} behaviour, in which the unbinding rate of a single dynein decreases with increasing force, for a certain range of force. Motivated by these experiments, we propose a model for catch bonding in dynein using a threshold force bond deformation (TFBD) model wherein catch bonding sets in beyond a critical applied load force. We study the effect of catch bonding on unidirectional transport properties of cellular cargo carried by {\em multiple} dynein motors within the framework of this model. We find catch bonding can result in dramatic changes in the transport properties, which are in sharp contrast to kinesin driven unidirectional transport, where catch bonding is absent. We predict that, under certain conditions, the average velocity of the cellular cargo can actually increase as applied load is increased. We characterize the transport properties in terms of a velocity profile phase plot in the parameter space of the catch bond strength and the stall force of the motor. This phase plot yields predictions that may be experimentally accessed by suitable modifications of motor transport and binding properties. Our work necessitates a reexamination of existing theories of collective bidirectional transport of cellular cargo where the catch bond effect of dynein described in this Letter is expected to play a crucial role.
\end{abstract}

\pacs{87.16.A-}
\pacs{87.16.dj}
\pacs{87.16.Ka}
\pacs{87.16.Nn}

\date{\today}
\maketitle

Motor protein driven transport of cellular cargoes along polar microtubule (MT) filaments is one of the principal mechanisms by which active long distance transport is achieved within an eukaryotic cell \cite{cell,welte}. This mechanism plays a vital role in keeping the cell spatially organized and maintaining the uneven distributions of the various cellular components\cite{cell,welte}. While single motor properties has been extensively studied, both in experiments and theory \cite{howard}, a large class of cooperative transport processes depends critically on the interaction of various motors and their collective behaviour, which can give rise to a whole new class of emergent phenomena \cite{roop-cell,lipo-uni,lipo-bd1,roop-bidi-pnas, manoj,nature-hancock}. 

The mechanism of this cooperative transport, however, remains an important open question. 
Experiments have revealed that unidirectional transport of cellular cargo involves teamwork of motor proteins of a single type, e.g; kinesin, dynein and myosin family of motors, while bidirectional transport requires team of oppositely directed kinesin and dynein motors \cite{roop-cell, gelfand, welte}. While molecular architecture and transport properties of these different classes of motors are significantly diverse and different, existing theoretical studies, have  used the kinesin motor as a paradigm for motor-driven transport \cite{lipo-uni,lipo-bd1}. Crucially, the single motor unbinding rate is modeled as an exponentially increasing function of force (slip-bond) for both plus-end and minus-end directed motors \cite{lipo-uni,lipo-bd1}. However, recent experiments have shown that dynein unlike kinesin, exhibits {\it catch bond} behaviour, where beyond a certain threshold force, the detachment rate of a single dynein from MT filament decreases with increasing load force \cite{data-ambarish,roop-trends}.
  
Catch bond behaviour \cite{prezhdo-rev, mcever, rakshit} has been observed in various biological protein receptor-ligand complexes, such as the complex of leukocyte adhesion molecule P-selectin with the ligand PSGL-1 \cite{catchbond-ligand} and actin/myosin complex \cite{catchbond-actin, igna-myosin} as well as microtubule-kinetochore attachments \cite{akiyoshi}. Different mechanisms have been proposed for the catch bond such as the two-state two-pathway model \cite{evans, baresgov}, the one-state two-pathway model \cite{bartolo, pereverzev, novikova} and the bond deformation model \cite{prezhdo}. The one-state two-pathway model provides two dissociation pathways, a catch barrier which increases with force, and a slip barrier which decreases with force. If the catch barrier is initially lower than the slip barrier, the system demonstrates a catch-slip transition with increasing force. The deformation model, on the other hand, proposes that force alters the conformation space in a fashion that strengthens receptor-ligand binding, and hence decreases the detachment rate. If the minimum of the potential decreases faster with force than the height of the barrier, one again obtains a catch-slip behaviour \cite{prezhdo-rev}. 

\begin{figure}[] 
\begin{center}
\includegraphics[width=\linewidth]{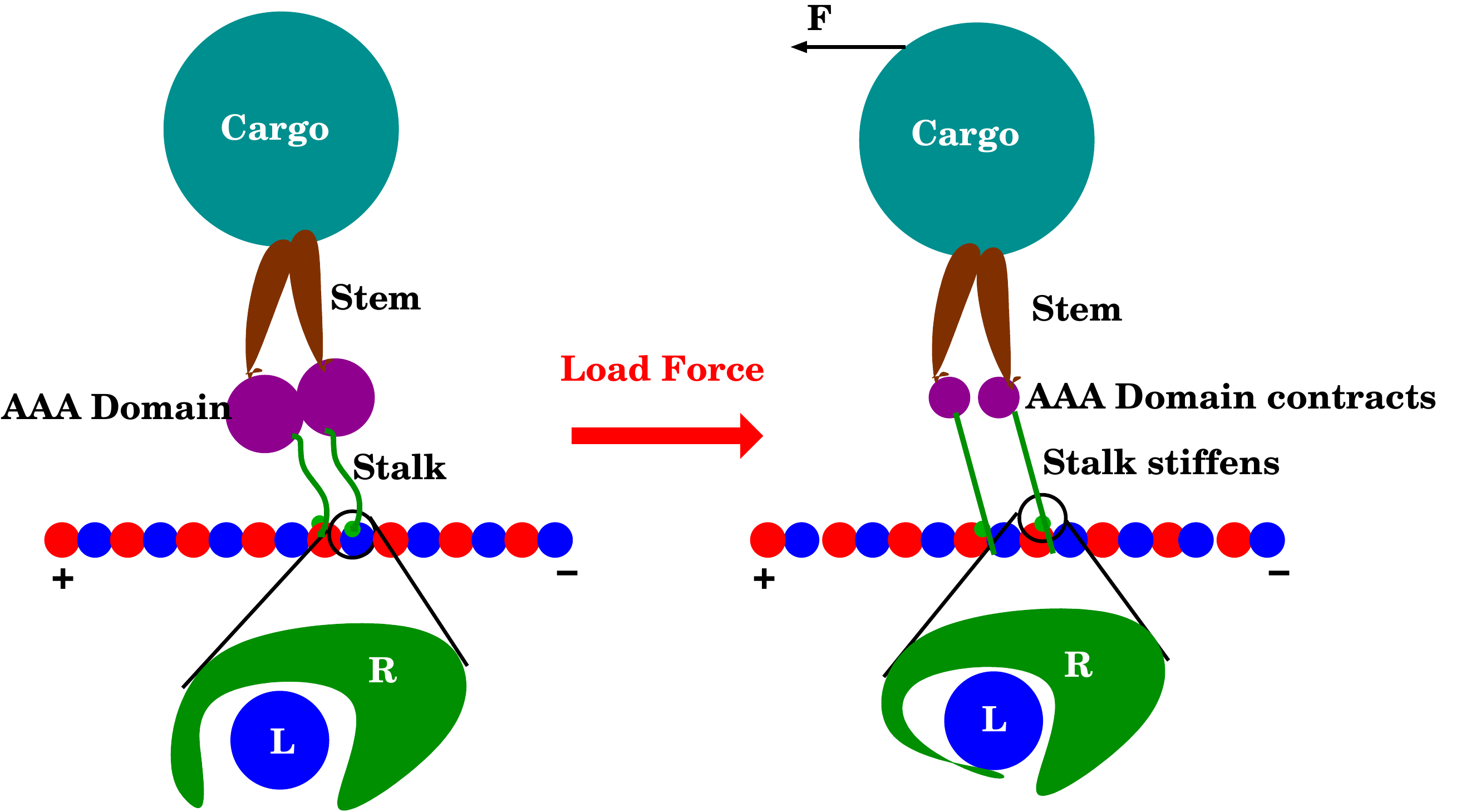} 
\caption{(Color online) Schematic representation of dynein walking on an MT filament, and catch bonding under applied load. The magnified region shows the MT binding domain of the dynein stalk, which can undergo a conformational change under applied load.}
\label{fig:schematic}
\end{center}
\end{figure}

In this work, motivated by recent experimental and simulation studies of dynein structure, we focus on the deformation model of catch bonds. Cytoplasmic dynein has two heads that walk processively along the microtubule stalk in discrete steps. Each head has a globular region consisting of six AAA domains \cite{roop-nature, roop-trends, roop-montecarlo}. This globular region has two elongated structures emerging from it, the stalk, which binds to the microtubule, and the stem, which binds to the cargo \cite{roop-trends}. It has been proposed that the globular head region contracts under applied load, which in turn causes tension to develop along the MT-binding stalk \cite{roop-nature, roop-trends}. Beyond a certain critical load, this can lead to allosteric deformations in the receptor region of the dynein stalk and the ligand domain on the MT surface which lock them together (Fig. \ref{fig:schematic}), resulting in a catch bond \cite{roop-trends}. At low/intermediate loads, the catch bond cannot be activated and differential stepping is required to advance against load \cite{roop-trends}. Direct experimental evidence comes from recent in-vitro experiments on single-molecule dynein detachment kinetics \cite{data-ambarish}. The detachment rate of a dynein motor is found to initially increase, and then decrease with force beyond a critical threshold. At extremely large forces, we should eventually regain a slip bond, thus exhibiting a slip-catch-clip behaviour over the entire force range. This we model through a Threshold Force Bond Deformation (TFBD) Model, where the deformation pathway activates beyond a certain threshold force. 


{\bf Model:} We consider a cargo which is transported on a filament by $N$ dynein motors against a constant external load force $F$. The state of the cargo is characterized by the number of bound motors $k$ ($0\leq k \le N$). Motors are irreversibly attached to the cargo but undergo attachment (detachment) to (from) the filament. With $k$ bound motors, the rates of attachment and detachment are given by $\pi_k = (N - k)\pi_{ad}$ and $\varepsilon_k = k\varepsilon$ respectively.
The load force is assumed to be shared equally by the $k$ bound motors \cite{lipo-uni}, so that each motor experiences a force $f = F/k$. The dynamics of the attachment/detachment process is given by the temporal evolution of the probability $p_k$ of having $k$ bound motors, expressed as the one-step master equation,
\begin{equation}
\frac{dp_k}{dt} = \varepsilon_{k+1}p_{n+1} + \pi_{k-1}p_{k-1} - (\varepsilon_k + \pi_k)p_k.
\label{eq:master}
\end{equation}
The corresponding master equation for the unbound vesicle $(k=0)$ is, $\frac{dp_0}{dt} = \varepsilon_{1}p_{1} - \pi_{0}p_{0}$, while for vesicle with $N$ bound motors, it is $\frac{dp_N}{dt} = \pi_{N-1}p_{N-1} - \varepsilon_{N}p_{N}$.

The bond deformation model proposes that catch bond behavior occurs by lowering the bound state due to force induced deformation of the bond \cite{prezhdo}. The deformation energy is given by,
$E_d(f) = \alpha [1 - \exp (-\frac{f}{f_0} ) ]$, 
where $\alpha$ characterizes the strength of the deformation energy and $f_0$ sets a force scale.

Since {\em in-vitro} dynein exhibits catch bond behaviour  above a threshold force  $f >  f_m$, and the load force $F$ is shared between $k$ motors, we introduce the Threshold Force Bond Deformation (TFBD) model with the deformation energy now given by,
\begin{equation}
E_d(f= F/k) = \Theta (f - f_m)\alpha \left[1 - \exp\left(-\frac{f - f_m}{f_0}\right)\right],
\end{equation}
and the unbinding rate of the cargo carried with $k$ attached motors attached to filaments is 
\begin{equation}
\varepsilon_{k}(f=F/k) = k\varepsilon_0\exp\left[ - E_d(f) + f/f_d\right]
\label{eq:tfbd}
\end{equation}
where the second term represents the usual slip contribution which exponentially grows with applied load. This TFBD Model exhibits a {\it slip-catch-slip} behavior for a single motor unbinding rate as a function of applied force on the motor. In a more general context, we also study the Bond Deformation (BD) model, by setting $f_m$ to zero in Eq.(~\ref{eq:tfbd}).

A cargo which is bound by $k$ motors moves with a velocity $v_k$. With increasing load force, the velocity of the cargo is expected to decrease until it comes to a rest at some critical stall force $f_s$. The decrease is approximately linear, as has been measured for kinesin \cite{howard, kinesin-stall} and cytoplasmic dynein \cite{roop-dynein-stall, lipo-bd2}. This is modelled by the following force velocity relation
\begin{equation}
v_k = v_0\left[ 1 - \left(\frac{F}{k f_s}\right) \right]
\end{equation}
where $v_0$ is the zero-force velocity for $k$ bound motors which is assumed to be independent of the number of bound motors. 

The steady state solutions to the master equation yield the probabilities for the unbound and the various bound motor states. The different transport properties are obtained after normalizing the probabilities with respect to the bound motor states alone \cite{lipo-uni}. 

\begin{figure}[t!!] 
\begin{center}
\includegraphics[width= 2.6 in,angle=270]{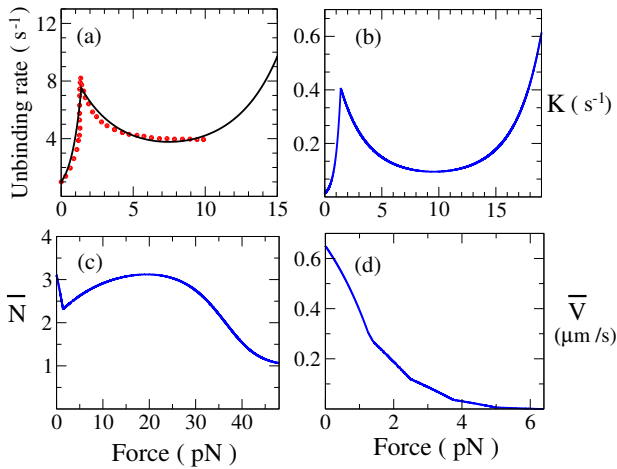}
\caption{(Color online) (a) Variation of unbinding rate of a {\it single} dynein with constant load force (F): The circles correspond to data points obtained from Ref.\cite{data-ambarish}. The solid curve is obtained from TFBD model with $\alpha/k_{B}T = 68$, $f_{o} = 40.7 pN$, $f_{m} = 1.4 pN$, $f_d = 0.67 pN$ with  $f_s = 1.25 pN$ \cite{data-ambarish} and $\varepsilon_0 = 1 s^{-1}$ \cite{data-ambarish}. Transport properties of cargo carried by 5 dyneins as function of load force (F):(b) Effective Unbinding rate (K) vs $F$. (c) The average number of attached motors vs F, (d) Average velocity vs F. For (b), (c) and (d), parameter values chosen are same as (a) with $\pi_{ad} = 1.6 s^{-1}$\cite{lipo-bd1}, $v_{o} = 0.65 \mu m/s$\cite{lipo-bd1}.}
\label{fig:tfbd}
\end{center}
\end{figure}


\begin{figure}[t!] 
\begin{center}
\includegraphics[width=\linewidth]{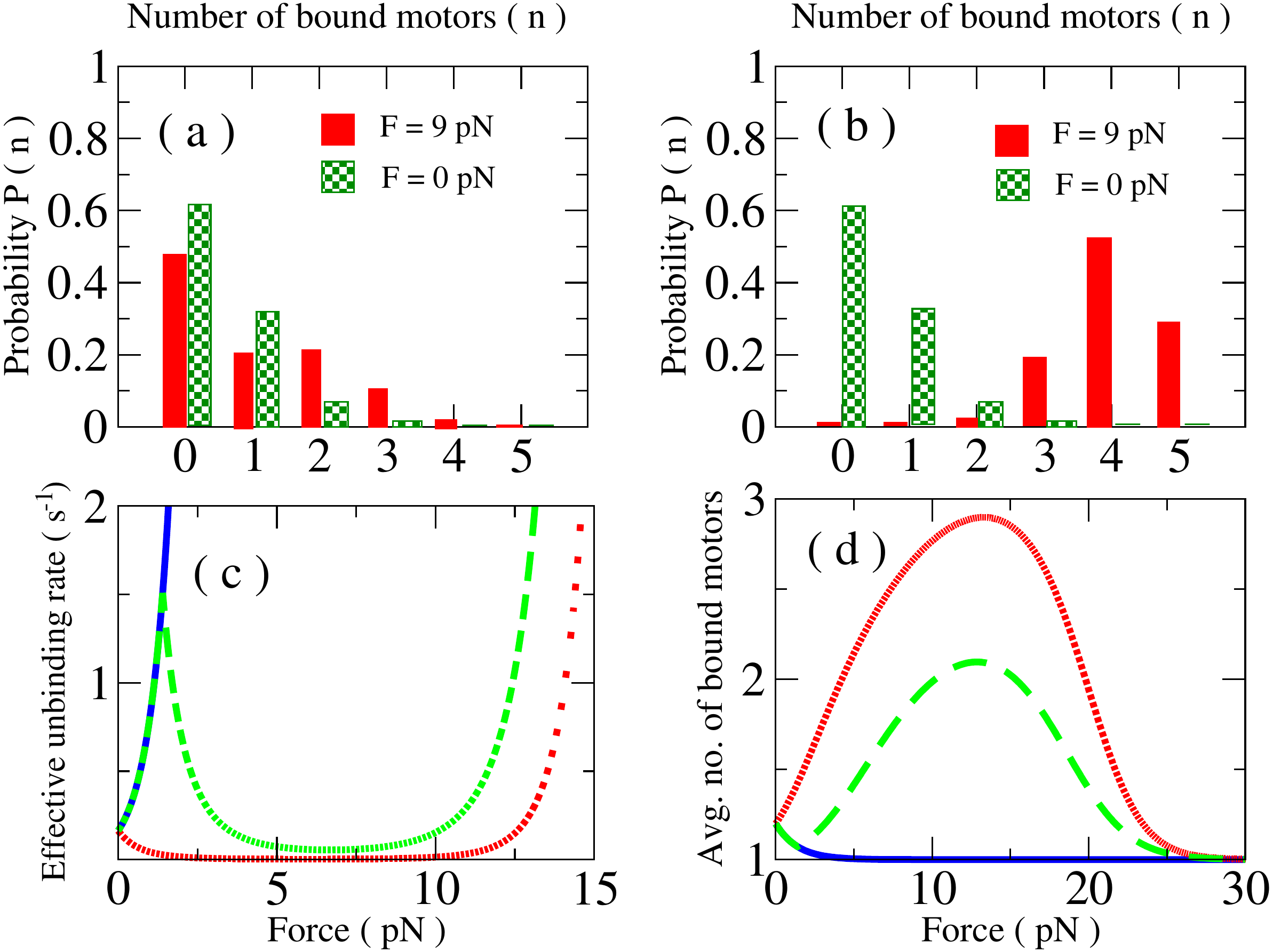} 
\caption{(Color online) Probability distribution for the number of bound motors in a system with $N=5$ for the (a) TFBD and (b) BD model. Compared to the zero force case (green crosses), where the probability distribution is peaked at zero bound motors and then decreases monotonically, for finite forces, both the TFBD and the BD model show a non-monotonic probability distribution, with the distribution peaking at larger $n$ in the BD case. Panels (c) and (d) shows the effective unbinding rate and the average number of bound motors as a function of force for a slip bond (solid blue line), as opposed to the TFBD (green dashed curve) and the BD model (red dotted line). Data is for
$\alpha/k_{B}T = 20 $ and $f_s = 2 pN$ with $v_0 = 0.65 \mu m/s$ \cite{lipo-bd1} , $\varepsilon_0 = 1.0 s^{-1}$ \cite{data-ambarish}, $f_0 = 7 pN, f_m = 1.4 pN, f_d = 0.67 pN$ and $\pi_{ad} = 0.1 s^{-1}$. }
\label{fig:prob-eps-n}
\end{center}
\end{figure}

{\bf Results:} Having described the model, we first fit the experimental data \cite{data-ambarish} of unbinding rate of single dynein from the MT filament with our proposed TFBD model. This model is able to capture the essential functional behaviour of unbinding rate of a single dynein as function of applied load force that has been observed in experiments, where the unbinding rate is seen to initially increase with increasing load starting with zero load, and subsequently decrease with increasing load beyond stall force $f_s$ as shown in Fig.~\ref{fig:tfbd}(a) \cite{data-ambarish}. In fact, we predict that the single motor unbinding rates should eventually increase with increased load force for forces higher than those accessed in this set of experimental studies. Next we study how the catch bonding behaviour exhibited by single dynein for unidirectional transport affects the steady state transport properties of cellular cargo that is being transported by {\it multiple} dynein motors. We consider a cellular cargo being carried by 5 dynein motors \cite{footnote}. We use the set of fitted model parameter values of the aforementioned experiment, to find the functional behaviour of various transport properties as a function of applied load force $F$. Fig.~\ref{fig:tfbd}(b) shows the effective unbinding rate of the cargo from the MT filament  increases initially and then it decreases for forces upto about 10 $pN$, following which it again starts increasing with increasing load force. Fig.~\ref{fig:tfbd}(c) shows that the average number of attached motors also increases after an initial dip and  finally for forces large than about 20 pN, the average number of attached motors decreases, and  exponentially approaches 1. Fig.~\ref{fig:tfbd}(d) shows that the mean velocity of the cargo decreases monotonically with increasing load in this parameter regime. 


\begin{figure}[t] 
\begin{center}
\includegraphics[width=0.9\linewidth]{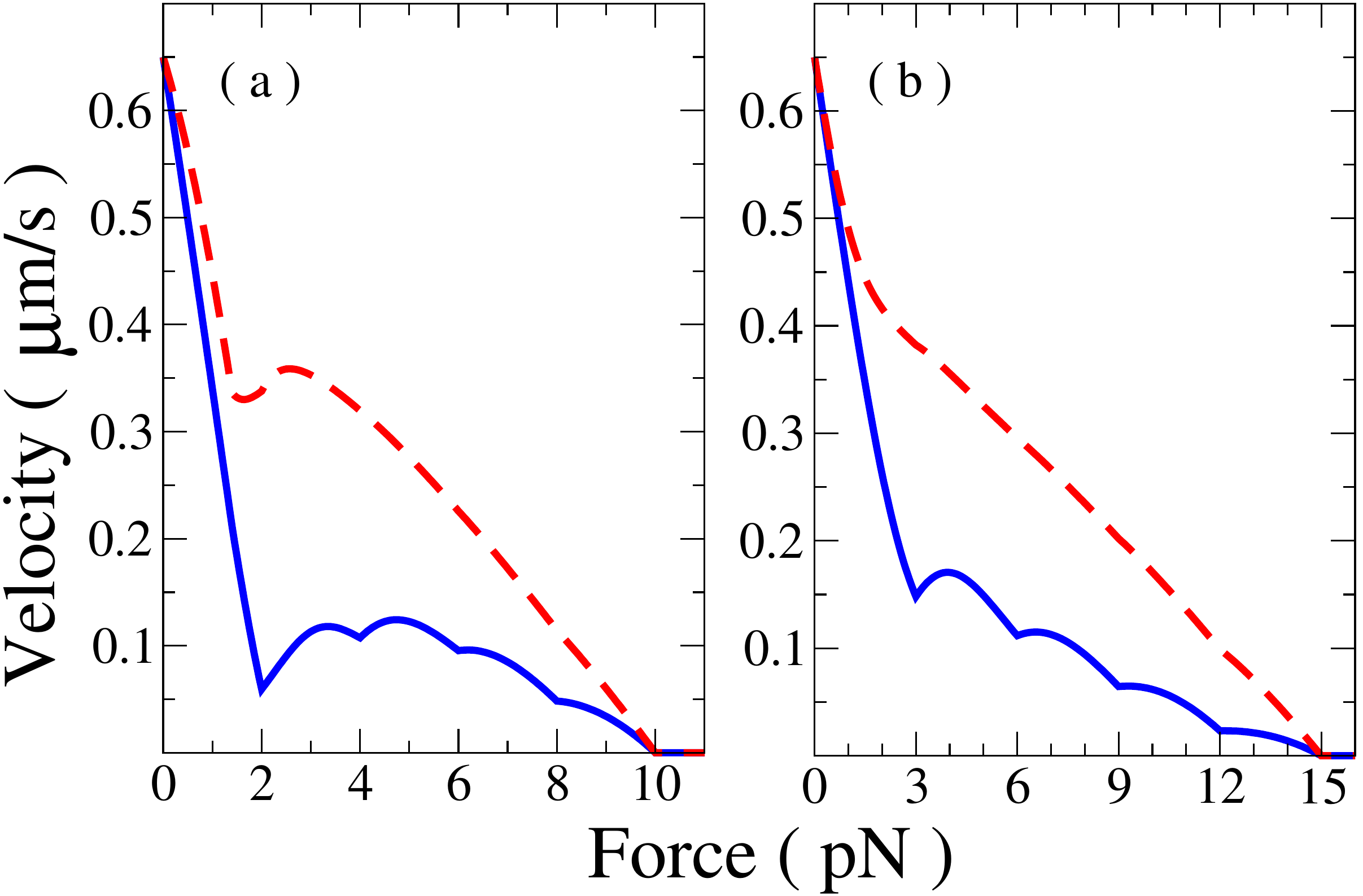} 
\caption{(Color online) Force-velocity curves: (a) TFBD Model for two different binding rates, $\pi_{ad} = 0.1 s^{-1}$ (solid blue curve) and $\pi_{ad} = 1.0 s^{-1}$ (dashed red curve). The solid curve shows four velocity humps which reduces to a single-hump velocity profile on increasing $\pi_{ad}$. (b) BD Model for $\pi_{ad} = 0.01 s^{-1}$ (solid blue curve) and $\pi_{ad} = 0.1 s^{-1}$ (dashed red curve). In this case, a four-hump profile reduces to a monotonically decreasing velocity profile on increasing $\pi_{ad}$. Data is for $N = 5$ motors with $v_0 = 0.65 \mu m/s$ \cite{lipo-bd1}, $\epsilon_0 = 1.0 s^{-1}$\cite{data-ambarish}, $f_0 = 7 pN, f_m = 1.4 pN, f_d = 0.67 pN, \alpha/k_{B}T = 35$ and $f_s = 2 pN$.}
\label{fig:velocity}
\end{center}
\end{figure}

Experimentally, it is known that the various motor properties can vary significantly for different classes of dynein motors. For instance, {\it weak} dynein is known to have a stall force of $f_s = 1.1 pN$ while for {\it strong} dynein $f_s = 7 pN$\cite{lipo-bd2}. We next explore the different {\it plausible} scenarios resulting from the  ramifications of generic catch bond behaviour observed not only for dynein motors but also for myosin motors \cite{catchbond-actin,igna-myosin} by studying the transport properties both for the TFBD model as well as the Bond Deformation (BD) model, by setting $f_m$ to zero in Eq.~\ref{eq:tfbd}. We shall focus on the variation of the transport properties resulting from changes in stall force $f_s$ , binding rates $\pi_{ad}$ and catch bond strength $\alpha$.

The effect of the catch bonding on the probability distribution of $n$ bound motor state  $P(n)$ is to shift the peak value of the distribution towards higher number of bound motors for certain range of forces both for the TFBD model  (Fig.~\ref{fig:prob-eps-n}a) and the BD model (Fig.~\ref{fig:prob-eps-n}b). This shift results from the fact that when a larger number of motors is bound to the filament, the force on each motor is low enough that they are in the {\it catch} regime, resulting in a decrease of propensity of individual motors to detach from filament in this state and hence a consequent increase in the probability of the states with higher number of attached motors. Conversely, when fewer number of motors bind to the filament, the load on each motor is higher, so that motors in this state are more likely to detach, and hence have a lower probability. This shifting of peak of the probability distribution towards  higher $n$ states manifests as an increase in the average number of bound motors and a decrease in effective unbinding rates for certain range of forces for both TFBD and BD model. This is shown in Figs.~\ref{fig:prob-eps-n}(c) and ~\ref{fig:prob-eps-n}(d) and the behaviour contrasts with the case of slip bonds where the effective unbinding rate monotonically increases and the average number of bound motors monotonically decreases with increasing load. 


Next we consider the mean velocity profile of the cargo as a function of load force. Catch bond behaviour manifests in rather remarkable behaviour for the velocity profile of the cargo - the mean velocity of the cargo can actually increase with increase in opposing load force for certain range of $f_s$, the strength of catch bond $\alpha$ and binding rate $\pi_{ad}$ for both the TFBD and BD model (Fig.~\ref{fig:velocity}). This unique behaviour can be understood as a direct consequence of the catch bond effect, which tends to stabilize the bound motor states with higher attached motors which move with a higher velocity so that the average velocity of the cargo actually increases.

\begin{figure}[t]
\centering
\includegraphics[width=0.99\linewidth]{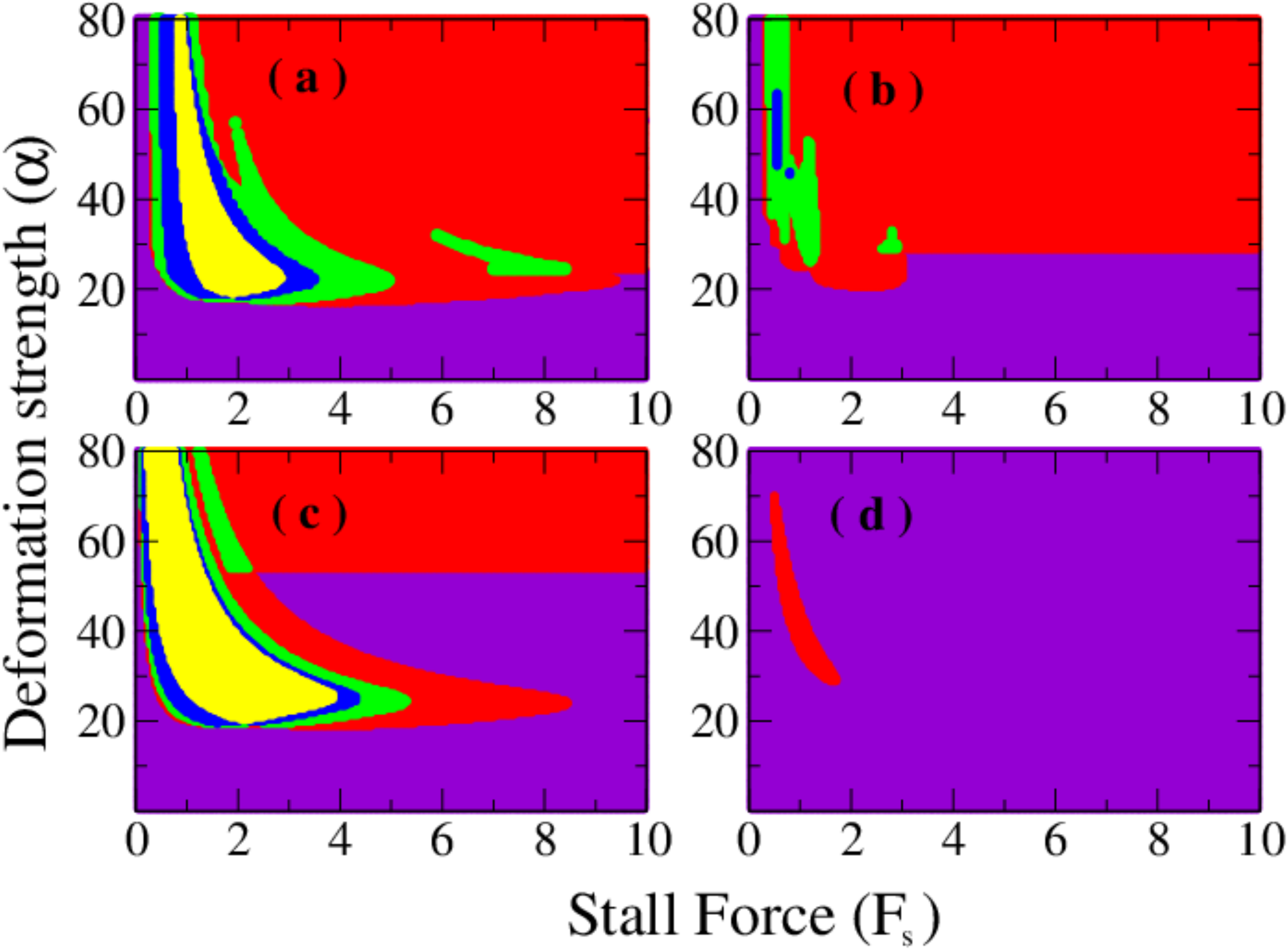}
\begin{center}
\caption{Velocity phase diagrams in the $\alpha - F_s$ plane for the TFBD model with $\pi_{ad} =$ (a) $0.1 s^{-1}$, (b) $1.0 s^{-1}$, and for the BD model with $\pi_{ad} =$ (a) $0.01 s^{-1}$, (b) $0.1 s^{-1}$. The violet region corresponds to a monotonically decreasing force velocity profile. The red, green blue and yellow regions corresponds to velocity profiles with one, two, three and four humps respectively (see Fig.~\ref{fig:velocity} for examples of individual velocity profiles). Parameter values are the same as in Fig.~\ref{fig:velocity}. Stall forces is in pNs. $\alpha$ is in in units of $k_{b} T$.}
\label{fig:phaseplot}
\end{center}
\end{figure}

We systematically study the effect of variation of the stall force $f_s$ and $\alpha$ by constructing a velocity profile phase diagram with different regions characterised by the number of maxima of mean velocity in the force-velocity profiles. Fig.~\ref{fig:phaseplot} shows the resulting phase diagrams for two different $\pi_{ad}$ values each for the TFBD (\ref{fig:phaseplot}(a) \ref{fig:phaseplot}(b)) and the BD (\ref{fig:phaseplot}(c) \ref{fig:phaseplot}(d)) models. The different regions in the phase diagram corresponds to different velocity-force profiles having maximas ranging from zero to 4 ( for N = 5). For sufficiently weak catch bond strength $\alpha$, the force-velocity  profiles is such that the mean velocity always decreases for increasing load force, similar to the behaviour in the absence of catch bond. However for both TFBD and BD model, increasing $\alpha$ and lowering $f_s$ have the effect of modifying the velocity-force profiles in a manner that they have one or more maxima of the mean velocity as illustrated in Fig.~\ref{fig:phaseplot}. The parameter space explored is a plausible biological regime and in principal should  be observable by suitable biochemical means which can alter and/or the stall forces, the catch bond strength and binding rates of the motors to the filament.

The TFBD Model proposed in this Letter captures the experimental results for dynein driven cargo transport. In contrast to canonical slip-bond models, the TFBD model for the catch bond has non-trivial consequences for the transport properties, in particular for the velocity profiles in response to applied loads, which should in principle be observable in experiments, and hence provides a testable prediction for our model. The more generic BD model also has similar dramatic differences in the transport properties that might be relevant for other motor driven systems. This work necessitates a reexamination of existing models of cellular cargo transport to take into account the catch bond mechanism described here. In particular, cooperative bi-directional cargo transport through the simultaneous action of oppositely directed motors (with one or both types of motors having a catch bond) is expected to have significantly different characteristics as compared to those described by existing theories, and will be discussed in a forthcoming publication. 

\section*{Acknowledgements}
MKM acknowledges financial support from the Ramanujan Fellowship, Department of Science and Technology, India and the IRCC Seed Grant, IIT Bombay. SM acknowledges DBT RGYI Project No: BT/PR6715/GBD/27/463/2012 for financial support. SC and AC acknowledges DST, India for financial support. The authors would also like to thank the organisers of the SMYIM conference, Pondicherry, where part of this work was done.


\begin{thebibliography}{}
\bibitem{footnote} It has been reported in Ref.\cite{roop-bidi-pnas} that for endosome transport four to eight dyneins are involved.
\end{thebibliography}
\end{document}